\documentclass[a4paper]{article}
\usepackage{epsfig,latexsym}

\newcommand{\bd}{\ensuremath{[B/D]}}
\newcommand{\var}[2][]{\mathrm{Var}_{#1}(#2)}
\newcommand{\cov}[3][]{\mathrm{Cov}_{#1}(#2,#3)}
\newcommand{\ex}[2][]{\mathrm{E}_{#1}(#2)}

\newcommand{\trans}{\mbox{}^\mathrm{T}}

\newcommand{\bvec}[1]{\mathbf{#1}}
\newcommand{\reduce}{\textsf{REDUCE}}
\newcommand{\mupad}{\textsf{\textsl{MuPAD}}}
\newcommand{\be}{\begin{eqnarray}}
\newcommand{\ee}{\end{eqnarray}}
\newcommand{\beq}{\begin{equation}}
\newcommand{\eeq}{\end{equation}}
\newcommand{\comment}[1]{}
\newcommand{\eqref}[1]{(\ref{#1})}

\newtheorem{thm}{Theorem}
\newtheorem{lemma}{Lemma}

\newenvironment{proof}{\par\noindent\textsf{Proof}\par}{\hfill $\Box$\par\vspace{0.1in}\par}

\title{Bayes linear variance adjustment for time series}
\author{D.~J.~Wilkinson\thanks{E-mail:
    \texttt{d.j.wilkinson@durham.ac.uk}, WWW:
    \texttt{http://fourier.dur.ac.uk:8000/djw.html}  }}
\date{\today}

\begin{document}
\maketitle

\begin{abstract}
This paper exhibits quadratic products of
linear combinations of observables which identify the covariance
structure underlying  the
univariate locally linear time series dynamic linear model. The first-
and second-order moments for the joint distribution over these
observables are given, allowing Bayes
linear learning for the underlying covariance structure for the time
series model. An example is given which illustrates the methodology
and highlights the practical implications of the theory.
\end{abstract}


\section{Introduction}
\label{intro}
In \cite{djwgdlm}, new methodology is developed for the
revision of covariance structures underlying two-step invertible
dynamic linear models (DLMs). Two-step invertible DLMs are essentially
models without a trend component.
 Here, the locally
linear DLM will be discussed, which does have a trend component, and so is not
two-step invertible. 
Interest will be
focussed on univariate DLMs, since it is simpler to explain the theory
in the univariate context. However, a full covariance matrix approach
may be taken to the multivariate counterpart, thus generalising the
work in \cite{djwgdlm}. 
It will be
shown how one may learn about the three different kinds of variance
associated with the locally linear dynamic linear model, using partial
prior specification for certain aspects of the model. The theory will
be applied to the modelling of sales of a particular product from a
wholesale depot, and the importance of good estimation of all of the
variance components will be demonstrated in the context of this
example. 

\section{Variance modelling}
\subsection{The linear model}
First consider the model for the time series $\{X_1,X_2,\ldots\}$.
\be
X_t &=& M_t + Y_{1t}, \quad\forall t\geq 1\\
M_t &=& M_{t-1} + N_t + Y_{2t}, \quad\forall t\geq 2\\
N_t &=& N_{t-1} + Y_{3t}, \quad\forall t\geq 2
\ee
where beliefs about $M_1$ and $N_1$ are specified \textit{a priori},
and the collection of quantities $\{Y_{jk}|j=1,2,3,\ k\geq 1\}$ have
expectation zero, and are mutually uncorrelated. Further,
$\var{Y_{jk}}=v_j$ does not depend on $k$ for $j=1,2,3,\ k\geq
1$. Using the terminology of \cite{dlm}, this is a second-order
description of the univariate locally linear time series DLM.
 Here interest focusses on Bayes linear methods for
learning about the covariance structure underlying this
model. Explicitly, the desire is to make inferences about the variances
of $Y_{1t}$, $Y_{2t}$ and $Y_{3t}$, in order that we might revise
specifications for $v_1$, $v_2$ and $v_3$. 

\subsection{The quadratic model}
Form the unobservable vector time series
$\{\bvec{Z}_1,\bvec{Z}_2,\ldots\}$, where
$\bvec{Z}_t=(Y_{1t}^2,Y_{2t}^2,Y_{3t}^2)\trans,\ \forall t\geq 1$. 
This time series is judged to be a second-order exchangeable sequence (that is,
mean, variance and covariance specifications for the sequence remain invariant
under an arbitrary permutation of the sequence),
as described in
\cite{mgexchbel}. Using the second-order exchangeability
representation theorem \cite{mgexchbel}, we may decompose the vectors,
$\bvec{Z}_t$ in the following way.
\begin{equation}
\bvec{Z}_t = \bvec{V} + \bvec{S}_t,\quad \forall t\geq 1
\label{eq:exchrep}
\end{equation}
where $\ex{\bvec{S}_t}=0,\ \cov{\bvec{V}}{\bvec{S}_t}=0,\ \forall t$ and
$\cov{\bvec{S}_s}{\bvec{S}_t}=0\ \forall s\not=t$.
The additional assumptions, $\cov{V_i}{V_j}=0,\ \forall i\not= j$ and
$\cov{S_{it}}{S_{jt}}=0\ \forall i\not=j,\forall t$, where
$\bvec{V}=(V_1,V_2,V_3)\trans$ and
$\bvec{S}_t=(S_{1t},S_{2t},S_{3t}),\forall t$, are made.
Note that the components of $\bvec{V}$ represent the underlying
variances for $Y_{1t}$, $Y_{2t}$ and $Y_{3t}$, and in particular, that
$\ex{\bvec{V}} =
(\var{Y_{1t}},\var{Y_{2t}},\var{Y_{3t}})\trans$. Revised beliefs about $\bvec{V}$
will lead to revised specifications for $v_1$, $v_2$ and $v_3$. Next, observables are
constructed which are predictive for $\bvec{V}$.

\section{State independent observables}
\subsection{Linear observables}
\label{sec:foo}
First construct the one-step differenced time series,
$\{X_2^\prime,X_3^\prime,\ldots\}$, where
\be
X_t^\prime &=& X_t - X_{t-1}\\
&=& N_t + Y_{2t} + Y_{1t} - Y_{1(t-1)}, \quad\forall t\geq 2
\ee
Next construct the one-, two-, and three-step differences of the
differenced series, $\{X^{(1)}_3,X^{(1)}_4,\ldots\}$,
$\{X^{(2)}_4,X^{(2)}_5,\ldots\}$ and 
$\{X^{(3)}_5,X^{(3)}_6,\ldots\}$, where
\be
X^{(1)}_t &=& X^\prime_t - X^\prime_{t-1}\nonumber \\
&=& Y_{3t} + Y_{2t}- Y_{2(t-1)} + Y_{1t} - 2Y_{1(t-1)}  +
Y_{1(t-2)},\quad \forall 
t\geq 3 \label{eq:x1}\\
X^{(2)}_t &=& X^\prime_t - X^\prime_{t-2} \nonumber\\
&=&   Y_{3t} + Y_{3(t-1)}+ Y_{2t} - Y_{2(t-2)}+Y_{1t} - Y_{1(t-1)}
-Y_{1(t-2)} + 
Y_{1(t-3)},\quad \forall t\geq 4 \label{eq:x2}\\
X^{(3)}_t &=& X^\prime_t - X^\prime_{t-3}\nonumber \\
&=&  Y_{3t}+ Y_{3(t-1)}  + Y_{3(t-2)} + Y_{2t} -
Y_{2(t-3)}+Y_{1t}- Y_{1(t-1)} -Y_{1(t-3)} + Y_{1(t-4)},\nonumber\\
&&\hspace{4in} \forall t\geq 5 \label{eq:x3}
\ee
Note that these observable series only involve the error
structure. Also note that the $\{X^{(1)}_3,X^{(1)}_4,\ldots\}$ series
is (second-order) $3$-step exchangeable, as defined in
\cite{djwgdlm} and discussed more fully in
\cite{djwthesis}.
Briefly, a collection of random quantities is (second-order) $n$-step
exchangeable 
if the expectation and covariance structure over them is invariant
under a reflection 
or arbitrary translation of the collection, and if the covariance
between any two members of the collection is fixed provided only that
they are a distance of at least $n$ apart.
Note similarly that the $\{X^{(2)}_4,X^{(2)}_5,\ldots\}$ series is
$4$-step exchangeable and that the $\{X^{(3)}_5,X^{(3)}_6,\ldots\}$
series is $5$-step exchangeable.

\subsection{Quadratic observables}
\label{sec:bar}
Form the series of the squares of the linear series, $\{{X^{(1)}_3}^2,{X^{(1)}_4}^2,\ldots\}$,
$\{{X^{(2)}_4}^2,{X^{(2)}_5}^2,\ldots\}$ and 
$\{{X^{(3)}_5}^2,{X^{(3)}_6}^2,\ldots\}$. Note that due to assumptions
of (second-order) exchangeability for the quadratic residuals, these
series have 
the same $n$-step exchangeability properties as the linear series
they are constructed from. Consequently, the (second-order) $n$-step
exchangeability representation theorem  \cite{djwgdlm} tells us that
the series \emph{identify} the Cauchy limit of the partial arithmetic
means. A random quantity is identified by a collection of observables if
as much uncertainty as is desired may be resolved by observing an increasing 
number of the observables.
\begin{lemma}
The following identification results hold.
\begin{itemize}
\item $\{{X^{(1)}_t}^2|\forall t\geq 3\}$ identify $6V_1 + 2V_2 +
  V_3$
\item $\{{X^{(2)}_t}^2|\forall t\geq 4\}$ identify $4V_1 + 2V_2 +
  2V_3$
\item $\{{X^{(3)}_t}^2|\forall t\geq 5\}$ identify $4V_1 + 2V_2 +
  3V_3$
\end{itemize}
\label{lem:id}
\end{lemma}
\begin{proof}
Using \eqref{eq:exchrep} and \eqref{eq:x1}, the collection
$\{{X^{(1)}_t}^2|\forall t\geq 3\}$ identifies
\be
\lim_{N\rightarrow\infty}\frac{1}{N}\sum_{t=3}^{N} {X^{(1)}_t}^2 &=&
\lim_{N\rightarrow\infty}\frac{1}{N}\sum_{t=3}^{N}
(Y_{3t}+Y_{2t}+Y_{1t}-2Y_{1(t-1)}-Y_{2(t-1)}+Y_{1(t-2)})^2\\
&=& \lim_{N\rightarrow\infty}\frac{1}{N}\sum_{t=3}^{N}
(Y_{3t}^2+Y_{2t}^2+Y_{1t}^2+4Y_{1(t-1)}^2+Y_{2(t-1)}^2+Y_{1(t-2)}^2)\\
&=& V_3+V_2+V_1+4V_1+V_2+V_1\\
&=& 6V_1 + 2V_2 + V_3
\ee
Using \eqref{eq:x2} and \eqref{eq:x3}, the other results follow
similarly.   
\end{proof}
In fact, $\{ {X_t^{(n)}}^2|\forall t\geq n+2 \}$ identifies
$4V_1+2V_2+nV_3,\ \forall n\geq 2$.
Note also  that
\begin{equation}
\left(\begin{array}{ccc}
6&2&1\\
4&2&2\\
4&2&3
\end{array}\right)^{-1} =
\left(\begin{array}{ccc}
1/2&-1&1/2\\
-1&7/2&-2\\
0&-1&1
\end{array}\right)
\label{eq:inv}
\end{equation}
and so Lemma \ref{lem:id} inverts to give
\begin{thm}
The following identification results hold.
\begin{itemize}
\item $\{\frac{1}{2}{X^{(1)}_t}^2-{X^{(2)}_t}^2+\frac{1}{2}{X^{(3)}_t}^2
|\forall t\geq 5\}$ identify $V_1$
\item $\{-{X^{(1)}_t}^2+\frac{7}{2}{X^{(2)}_t}^2-2{X^{(3)}_t}^2
|\forall t\geq 5\}$ identify $V_2$
\item $\{-{X^{(2)}_t}^2+{X^{(3)}_t}^2
|\forall t\geq 5\}$ identify $V_3$
\end{itemize}
\label{thm:id}
\end{thm}
\begin{proof}
Clear from Lemma \ref{lem:id} and \eqref{eq:inv}.
\end{proof}
Of course, the partial arithmetic means of these collections may be
used as frequentist unbiased estimators of the underlying variances (though
they would not necessarily have optimum variance properties). Here,
the quadratic observables will be used in order to allow Bayes linear
updating of beliefs for the underlying variances.

\section{Bayes linear methods}
A Bayes linear approach is taken to subjective statistical inference,
making expectation (rather than probability) primitive. An overview of
the methodology is given in \cite{fgcross}.
 The emphasis of
this paper is on learning about underlying means; however, 
 the foundations of the theory are quite general, and are outlined in
the context of second-order exchangeability in \cite{mgrevexch},
and discussed for more general situations in \cite{mgpriorinf}.
Bayes
linear methods may be used in order to learn about any quantities of
interest, provided only that a mean and variance specification is made
for all relevant quantities, and a specification for the covariance
between all pairs of quantities is made. No distributional assumptions
are necessary.
There are many interpretive and
diagnostic features of the Bayes linear methodology. These are
discussed with reference to \bd\ (the computer language used for the analysis of
the example given in this paper) in \cite{gwblincomp}. 

\section{Covariance structure over the quadratic observables}
In order to carry out Bayes linear updating of the underlying
variances, the covariances over and between the underlying variables and the
predictive observables are required. They are given as follows:
\begin{lemma}
\be
\cov{V_1}{{X_t^{(1)}}^2} &=& 6\var{V_1},\quad \forall t\geq 3\\
\cov{V_1}{{X_t^{(2)}}^2} &=& 4\var{V_1},\quad \forall t\geq 4\\
\cov{V_1}{{X_t^{(3)}}^2} &=& 4\var{V_1},\quad \forall t\geq 5\\
\cov{V_2}{{X_t^{(1)}}^2} &=& 2\var{V_2},\quad \forall t\geq 3\\
\cov{V_2}{{X_t^{(2)}}^2} &=& 2\var{V_2},\quad \forall t\geq 4\\
\cov{V_2}{{X_t^{(3)}}^2} &=& 2\var{V_2},\quad \forall t\geq 5\\
\cov{V_3}{{X_t^{(1)}}^2} &=& 1\var{V_3},\quad \forall t\geq 3\\
\cov{V_3}{{X_t^{(2)}}^2} &=& 2\var{V_3},\quad \forall t\geq 4\\
\cov{V_3}{{X_t^{(3)}}^2} &=& 3\var{V_3},\quad \forall t\geq 5
\ee
\end{lemma}
\begin{proof}
These are a trivial consequence of Lemma \ref{lem:id} and the $n$-step
exchangeability representation theorem.
\end{proof}
The covariances between the quadratic observables themselves are
rather complex, and are given in the appendix.

\section{Bayes linear adjustment for the variances}
Theorem \ref{thm:id} shows that Bayes linear fitting of the underlying
variances, $V_1$, $V_2$, and $V_3$ on sufficiently many quadratic
observables will eventually resolve all uncertainty about these
quantities. Of course, in practice one will have only a finite number
of observations, $N$, with which to update beliefs about the
underlying variance structure. 
The \emph{adjusted expectation} for
$\bvec{V}$, $\ex[D]{\bvec{V}}$ given the three finite series of
quadratic observables, 
$\{X^{(1)}_t| 3\leq t\leq N\}$, $\{X^{(2)}_t| 4\leq
t\leq N\}$ and 
$\{X^{(3)}_t| 5\leq t\leq N\}$ takes the form
\beq
\ex[D]{\bvec{V}} = \ex{\bvec{V}} + \cov{\bvec{V}}{\bvec{D}}[\var{\bvec{D}}]^\dagger[\bvec{D}-\ex{\bvec{D}}]
\label{eq:aex}
\eeq
where $\bvec{D}$ is the $(3N-9)$-dimensional vector
\beq
\bvec{D} = \left({X^{(1)}_3}^2,\ldots,{X^{(1)}_N}^2,{X^{(2)}_4}^2,\ldots,{X^{(2)}_N}^2,{X^{(3)}_5}^2,\ldots,{X^{(3)}_N}^2\right)\trans
\eeq
and $[\var{\bvec{D}}]^\dagger$ denotes the Moore-Penrose generalised
inverse of $\var{\bvec{D}}$.
All necessary covariances are given in the previous section and the
appendix. Note that there is nothing particularly special about the choice of
$\bvec{D}$ other than the fact that it is one of the smallest 
choices of $\bvec{D}$ which will lead to
identification of the underlying variance structure. It could be enlarged by 
introducing terms of the form ${X^{(4)}_t}^2$ etc. This would lead to a richer
projection space, and hence more efficient estimates. However, one would have to
compute the covariance structure over the extra observables, and this exercise would
quickly become computationally unattractive. 

\section{Example}
\begin{figure}[thbp!]
\begin{center}
\epsfig{file=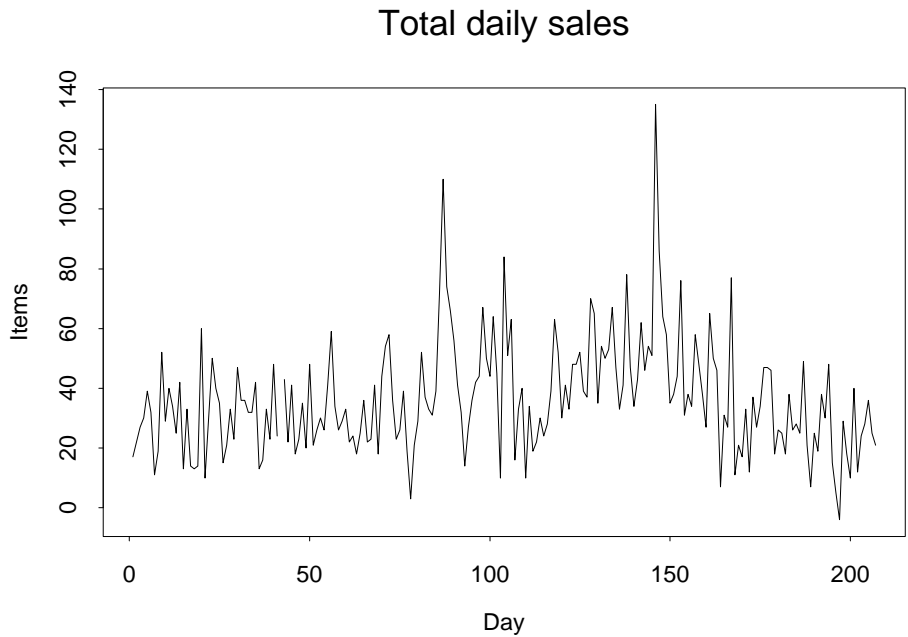,width=5in}
\end{center}
\caption{Plot showing the time series of sales for the example}
\label{fig:tsplot}
\end{figure}

\begin{figure}[thbp!]
\begin{center}
\epsfig{file=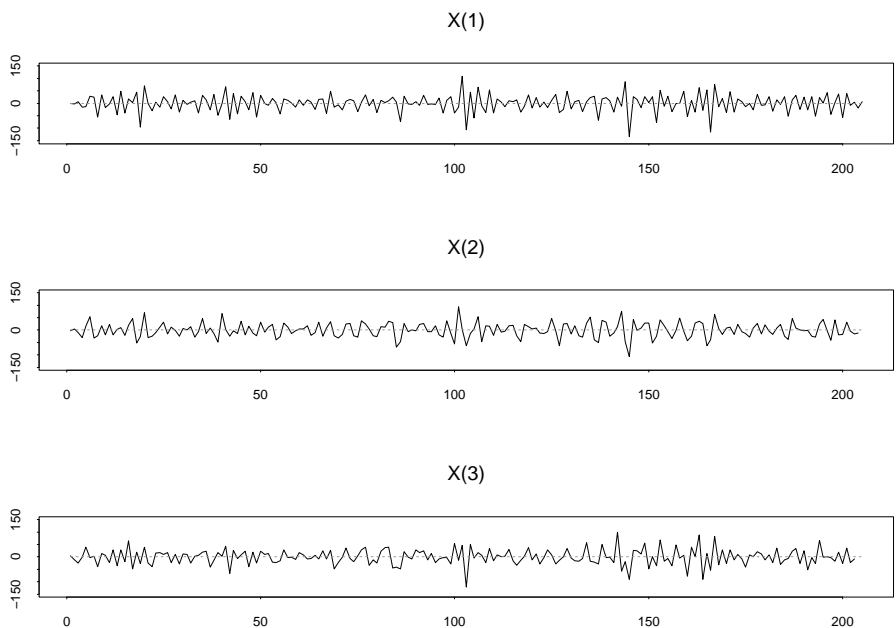,width=5in}
\end{center}
\caption{Plot showing the mean zero linear combinations}
\label{fig:linplot}
\end{figure}

The theory developed thus far will now be applied to a genuine example
arising from research into the forecasting of competitive retail
markets. Figure \ref{fig:tsplot} shows a time series for the case
sales of the leading brand of cola from a particular wholesale depot
in England for the first 200 days of 1995. It is assumed that these
sales figures, $\{X_t|1\leq t\leq 200\}$, follow a second-order
locally linear time-series DLM, 
and so the 3-, 4- and 5-step exchangeable sets of quantities,
$\{X^{(1)}_t| 3\leq t\leq 200\}$, $\{X^{(2)}_t| 4\leq
t\leq 200\}$ and 
$\{X^{(3)}_t| 5\leq t\leq 200\}$ are formed, as described in Section
\ref{sec:foo}, and these are shown in Figure
\ref{fig:linplot}. According to our model, these sets of quantities
are mean zero $n$-step exchangeable, and so obvious evidence of an
underlying mean away from zero, or evidence of long-range dependencies
would be evidence against the model. There does not appear to be any
obvious discrepancies. Note that short-range dependencies, and
dependencies between the series are to be expected.

\begin{figure}[thbp!]
\begin{center}
\epsfig{file=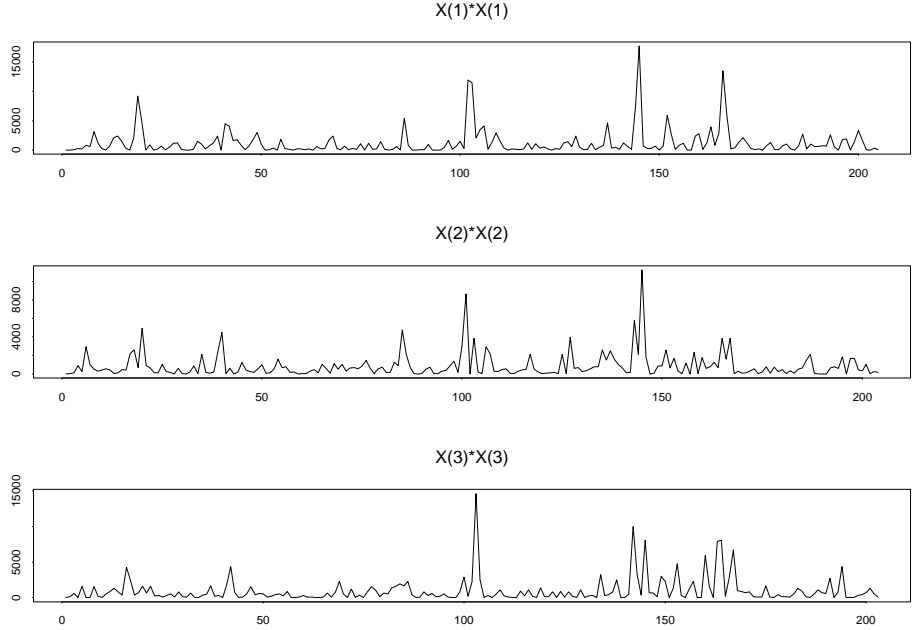,width=5in}
\end{center}
\caption{Plot showing quadratic observables}
\label{fig:quadplot}
\end{figure}

 Next,
the quadratic observables 
$\{{X^{(1)}_t}^2| 3\leq t\leq 200\}$, $\{{X^{(2)}_t}^2| 4\leq
t\leq 200\}$ and 
$\{{X^{(3)}_t}^2| 5\leq t\leq 200\}$ are formed, as described in Section
\ref{sec:bar}, and these are shown in Figure
\ref{fig:quadplot}. Again, due to assumptions about the
exchangeability of the quadratic residuals, these series are each $n$-step
exchangeable, and so obvious long-range dependencies would be evidence
against our model. Fortunately, none are apparent, although short
range dependencies, and dependencies between series are particularly
clear in this figure.

\textit{A priori} belief specifications are required before analysis
can take place. The specifications required for a basic linear
analysis of this problem were made as follows.
\be
\ex{M_1} &=& 20,\ \var{M_1}\ =\ 20^2,\ 
\ex{N_1} = 0,\ \var{N_1}\ =\ 3^2 \\
\ex{V_1}=\var{Y_{1t}} &=& 5^2,\ \ex{V_2}=\var{Y_{2t}}\ =\ 0.2^2,\ \ex{V_3}=\var{Y_{3t}}\ =\
0.1^2,\ \forall t
\ee
The above specifications are also precisely those required for a fully
Bayesian approach to the analysis of the locally linear time series
DLM, together with some distributional assumptions, as described
in \cite{dlm}. 
The additional specifications required for a quadratic analysis are
given below.
\be
\var{V_1} &=& 5^2,\ \var{V_2}\ =\ 1^2,\ \var{V_3}\ =\ 0.2^2\\
\var{S_{1t}} &=& 2(5^4),\ \var{S_{2t}}\ =\ 2(0.2^4),\ \var{S_{3t}}\
=\ 2(0.1^4),\ \forall t
\ee
Note that in this example, for simplicity, the variance specifications
for the $S_{it}$ have been made to be consistent with a $\chi^2$ fit
for the distribution of $Y_{it}^2$, given their underlying mean. Such
fitting is discussed in the more general multivariate context in
\cite{djwthesis}. Note however, that such fitting is not
required, and that in any case, assignment of the fourth moments in
this way is a much weaker assumption than that of full normality of
the $Y_{it}$.

Clearly, the additional specification burden
required in order to carry out variance learning is quite
small. Specification of six additional numbers (three, if one is
prepared to fit the $\var{S_{it}}$) is all that is required. Note also
that since we take a Bayes linear approach, adjustments reduce to the
solving of matrix equations, and so the computational requirements are
not great. 

\begin{figure}[thbp!]
\begin{center}
\epsfig{file=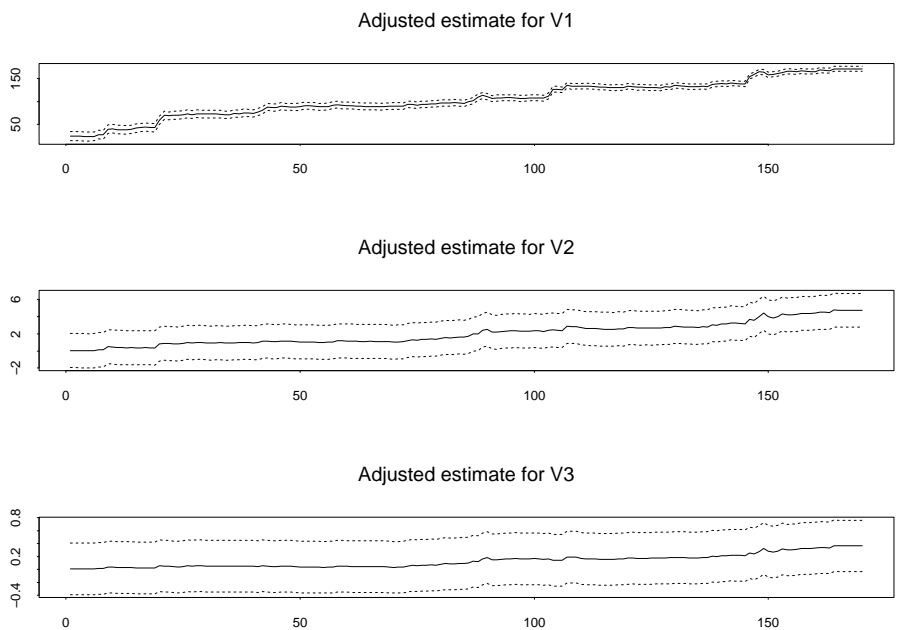,width=5in}
\end{center}
\caption{Variance component adjustments}
\label{fig:qas}
\end{figure}

\begin{figure}[thbp!]
\begin{center}
\epsfig{file=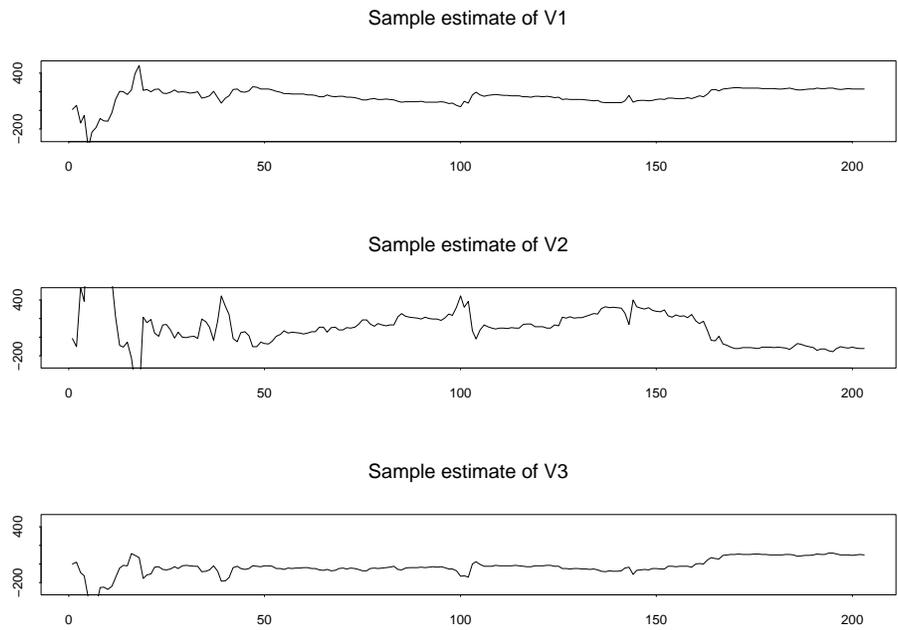,width=5in}
\end{center}
\caption{Unbiased sample estimates for the variance components}
\label{fig:unb}
\end{figure}

The Bayes linear computing package, \bd, was used to analyse the
problem. ``Elements'' corresponding to the linear and quadratic terms
were ``built'', and expectations and covariances were assigned
appropriately, using output from computer algebra systems where
necessary. The computer algebra systems were used for numerical
substitution of beliefs into the covariance formulae given in the
appendix, as well as for the algebraic derivation of the formulae
themselves.  Beliefs about $V_1$, $V_2$ and $V_3$ were then adjusted
using the quadratic observables formed from the first 200 observations
from the series. The sequence of adjusted expectations, together with
corresponding two-standard deviation credibility bounds for the
variance components are shown in Figure \ref{fig:qas}. Notice that
beliefs about all of the variances have been revised upwards. In
particular, beliefs  about the magnitude of $V_1$ have
been revised upwards quite sharply, relative to prior uncertainty. The
magnitude of  the
revisions should be regarded as a diagnostic warning for the high-order
variance specification, since the
eventual adjusted expectation lies some way outside the \emph{a
  priori} credibility 
bounds. The adjustments for the variances were as follows.
\beq
\ex[D]{V_1}=171,\quad \ex[D]{V_2}=4.75,\quad \ex[D]{V_3}=0.36
\eeq
 We shall see the implications of these revisions for first
order adjustments,  later. These adjustment sequences may be
compared with natural unbiased sample estimates for the variance
components, as shown in Figure \ref{fig:unb}. These estimates can be
seen to be unstable, due to the fact that the variance of these
estimates is very large, even with 200 observations.

\begin{figure}[thbp!]
\begin{center}
\epsfig{file=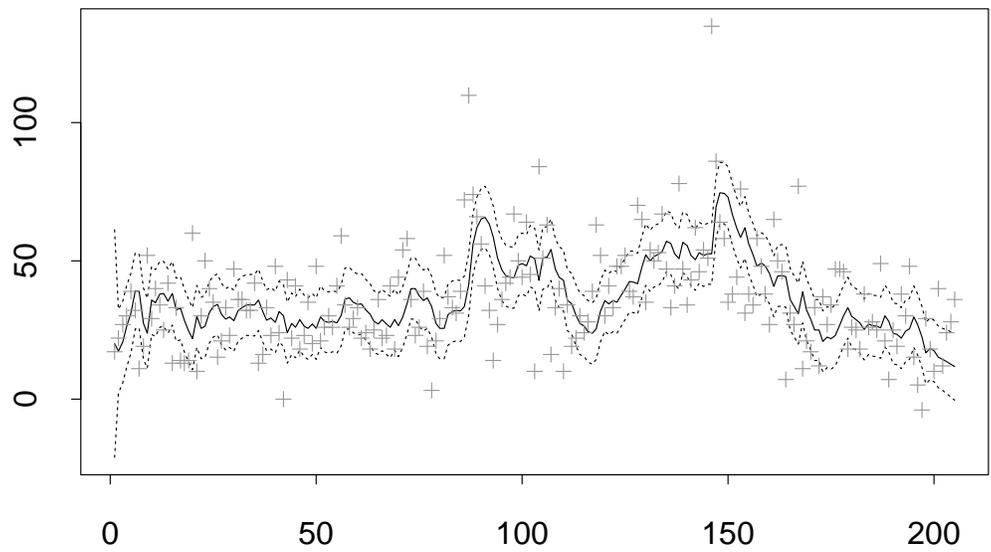,width=6in}
\end{center}
\caption{First order adjustments without variance revision}
\label{fig:foawovr}
\end{figure}

\begin{figure}[t!]
\begin{center}
\epsfig{file=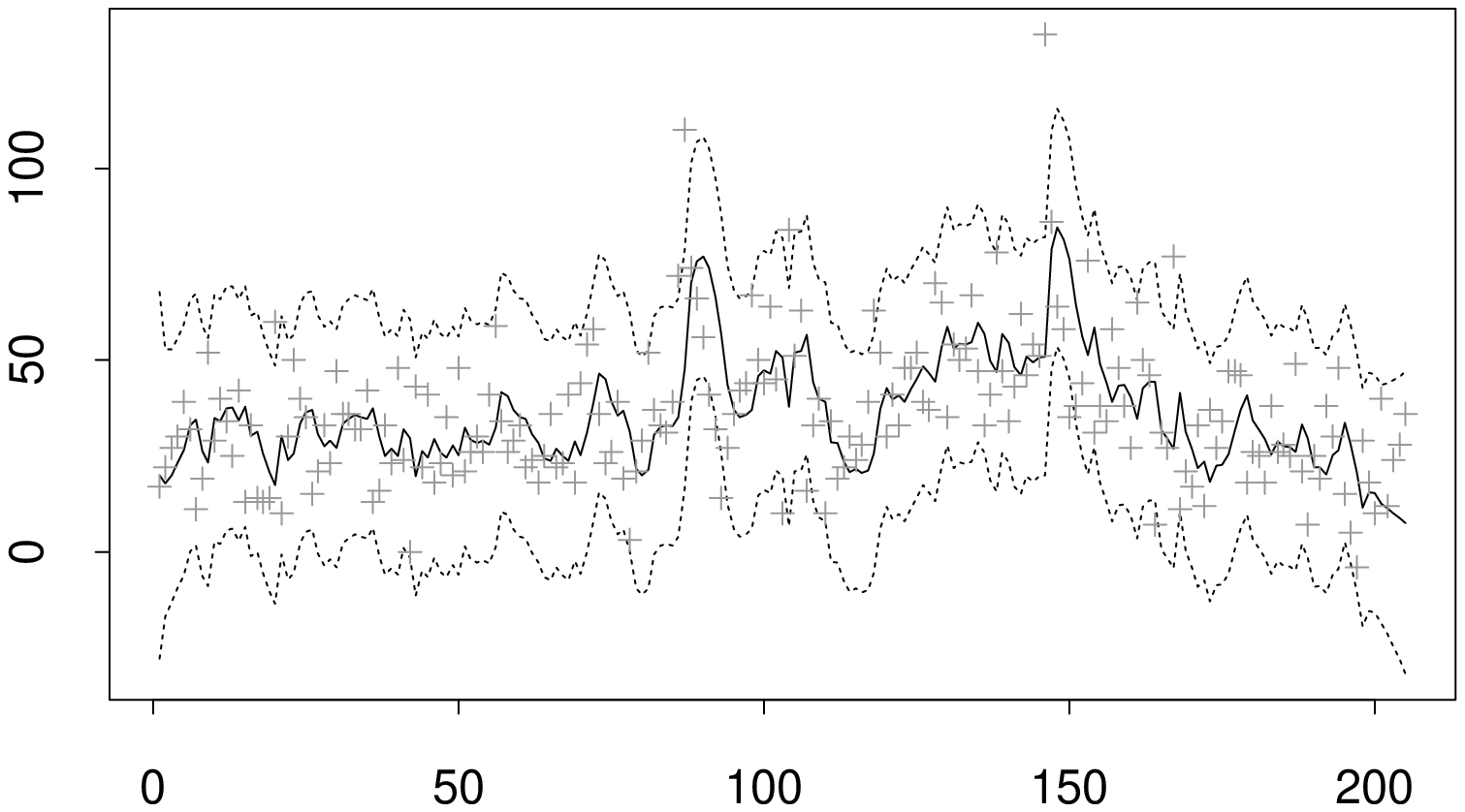,width=6in}
\end{center}
\caption{First order adjustments using revised variances}
\label{fig:foawvr}
\end{figure}

We may now consider the linear forecasting problem, and compare the
forecasts of that model using the original, and adjusted variance
specifications. Figure \ref{fig:foawovr} shows the first order adjustments,
together with a one-step ahead predictive two-standard deviation credibility
interval, using the original variance specifications. It appears that the 
specification for $\var{Y_{1t}}$ may be too small, since considerably more
than 5\% of observations lie outside the credibility interval. Figure 
\ref{fig:foawvr} shows the same, using the adjusted variances. The increased
specification for $\var{Y_{1t}}$ leads to wider credibility intervals, which
can be clearly seen to match better with the forecast performance of the
model. Notice also that the increased estimate for $V_2$ allows the
mean parameter, $M_t$, to adapt more quickly to fluctuations in the data.

\section{Conclusions}
Appropriate variance specifications are crucial to the performance of
dynamic linear models. Even a carefully chosen, and wholly appropriate
model will perform very poorly if the variance specifications used are wrong.
This fact has long been appreciated, and in \cite{dlm}, methods for
the updating of the top-level variance ($V_1$ in this paper) are discussed.
However, such methods remain the state of the art, and yet, not all
uncertainty for the top-level variance is resolved, and no methods are
given for data-driven learning for other variance components in the model.
Appropriate specifications for other variance components are just as
important, and can have just as much effect on the performance of the model
(especially for mean-dominated series, such as many financial time
series). In the example given in this paper, learning that $V_2$ was
bigger than thought allowed the series parameters to adapt more quickly
to the data. Such data-driven learning for the parameter variances is
not possible using existing DLM methodology. Further, it is hard to see how
the methodology might be adapted to allow such learning.
Here, methods are given for learning about all of the variance components
for a locally linear time series model, using limited partial prior 
specifications for the variance components, thus providing a subjective 
Bayesian, computationally tractable solution to the problem. 

\section{Acknowledgements}
All Bayes linear computations were carried out using the Bayes linear
computing package, \bd, available from \cite{wgbd}. The
covariance calculations were carried out using the \reduce\ computer
algebra system, described in \cite{reduce}, and the \mupad\
computer algebra system, described in \cite{mupadb} and
\cite{mupadp}. The data for the
example was provided by
\textit{Positive Concepts Ltd.}

\appendix

% \section*{Covariance structure over the quadratic observables}
\section*{Appendix}
\label{app:cov}
The covariances between the quadratic observables themselves are
rather complex, and were calculated using computer
algebra packages 
in order to ensure accuracy.

{\small
\begin{lemma}
The covariance structure over the ${X_t^{(1)}}^2$ terms is given by
the following relations.
\begin{eqnarray} 
\cov{{X_{t}^{(1)}}^2}{{X_{t}^{(1)}}^2} &=& \mathrm{Var}(S_{3t}) + 2\mathrm{Var}(S_{2t}) + 18\mathrm{Var}(S_{1t}) + \mathrm{Var}(V_{3}) +
 8\mathrm{Var}(V_{2})\nonumber\\
&& + 72\mathrm{Var}(V_{1})
 + 8\mathrm{E}(V_{3})\mathrm{E}(V_{2}) +
 24\mathrm{E}(V_{3})\mathrm{E}(V_{1}) + 4\mathrm{E}(V_{2})^2
 \nonumber\\
&&+ 48\mathrm{E}(V_{2})\mathrm{E}(V_{1}) +
36\mathrm{E}(V_{1})^2,\quad \forall t\geq 3\\
&&\nonumber\\ 
\cov{{X_{t}^{(1)}}^2}{{X_{t-1}^{(1)}}^2} &=& \mathrm{Var}(S_{2t}) + 8\mathrm{Var}(S_{1t}) + \mathrm{Var}(V_{3}) + 4\mathrm{Var}(V_{2}) + 52
\mathrm{Var}(V_{1})\nonumber\\
&& + 16\mathrm{E}(V_{2})\mathrm{E}(V_{1}) +
16\mathrm{E}(V_{1})^2,\quad \forall t\geq 4\\
&&\nonumber\\
\cov{{X_{t}^{(1)}}^2}{{X_{t-2}^{(1)}}^2} &=& \mathrm{Var}(S_{1t}) + \mathrm{Var}(V_{3}) +
4\mathrm{Var}(V_{2}) + 36\mathrm{Var}(V_{1}),\quad \forall t\geq 5\\
&&\nonumber\\ 
\cov{{X_{t}^{(1)}}^2}{{X_{t-s}^{(1)}}^2} &=& \mathrm{Var}(V_{3}) +
4\mathrm{Var}(V_{2}) + 36\mathrm{Var}(V_{1}),\quad \forall s\geq
3,\forall t\geq s+3 \label{eq:far1}
\end{eqnarray}
\end{lemma}

\begin{lemma}
The covariance structure over the ${X_t^{(2)}}^2$ terms is given by
the following relations.
\begin{eqnarray} 
\cov{{X_{t}^{(2)}}^2}{{X_{t}^{(2)}}^2} &=& 2(\mathrm{Var}(S_{3t}) + \mathrm{Var}(S_{2t}) + 2\mathrm{Var}(S_{1t}) + 4\mathrm{Var}(V_{3})
 + 4\mathrm{Var}(V_{2})\nonumber\\
&& + 20\mathrm{Var}(V_{1}) + 2\mathrm{E}(V_{3})^2 + 8\mathrm{E}(V_{3})
\mathrm{E}(V_{2}) + 16\mathrm{E}(V_{3})\mathrm{E}(V_{1}) +
2\mathrm{E}(V_{2})^2\nonumber\\ 
&& + 16\mathrm{E}(V_{2})\mathrm{E}(V_{1}) + 12\mathrm{E}(V_{1})^2),
\quad \forall t\geq 4\\
&&\nonumber\\
\cov{{X_{t}^{(2)}}^2}{{X_{t-1}^{(2)}}^2} &=& \mathrm{Var}(S_{3t}) +
3\mathrm{Var}(S_{1t}) + 
4\mathrm{Var}(V_{3}) + 4\mathrm{Var}(V_{2}) +  
12\mathrm{Var}(V_{1})\nonumber\\
&& - 4\mathrm{E}(V_{3})\mathrm{E}(V_{1}) -
4\mathrm{E}(V_{1})^2,
\quad \forall t\geq 5\\
&&\nonumber\\
\cov{{X_{t}^{(2)}}^2}{{X_{t-2}^{(2)}}^2} &=& \mathrm{Var}(S_{2t}) + 2\mathrm{Var}(S_{1t}) + 4\mathrm{Var}(V_{3}) + 4\mathrm{Var}(V_{2}) + 
20\mathrm{Var}(V_{1})\nonumber\\
&& + 8\mathrm{E}(V_{2})\mathrm{E}(V_{1}) +
4\mathrm{E}(V_{1})^2, \quad\forall t\geq 6\\
&&\nonumber\\
\cov{{X_{t}^{(2)}}^2}{{X_{t-3}^{(2)}}^2} &=& \mathrm{Var}(S_{1t}) +
4\mathrm{Var}(V_{3}) + 4\mathrm{Var}(V_{2}) + 16\mathrm{Var}(V_{1}),
\quad\forall t\geq 7\\
&&\nonumber\\
\cov{{X_{t}^{(2)}}^2}{{X_{t-s}^{(2)}}^2} &=& 4(\mathrm{Var}(V_{3}) +
\mathrm{Var}(V_{2}) + 4\mathrm{Var}(V_{1})), \quad\forall s\geq
4,\forall t\geq s+4 \label{eq:far2}
\end{eqnarray}
\end{lemma}

\begin{lemma}
The covariance structure over the ${X_t^{(3)}}^2$ terms is given by
the following relations.
\begin{eqnarray}
\cov{{X_{t}^{(3)}}^2}{{X_{t}^{(3)}}^2} &=& 3\mathrm{Var}(S_{3t}) + 2\mathrm{Var}(S_{2t}) + 4\mathrm{Var}(S_{1t}) + 21\mathrm{Var}(V(3
)) + 8\mathrm{Var}(V_{2})\nonumber\\
&& + 40\mathrm{Var}(V_{1}) + 12\mathrm{E}(V_{3})^2 +
24\mathrm{E}(V_{3})\mathrm{E}(V_{2})  +
48\mathrm{E}(V_{3})\mathrm{E}(V_{1}) + 4\mathrm{E}(V_{2})^2
\nonumber\\
&& + 32\mathrm{E}(V_{2})\mathrm{E}(V_{1}) + 24\mathrm{E}(V_{1})^2,
\quad\forall t\geq 5\\
&&\nonumber\\
\cov{{X_{t}^{(3)}}^2}{{X_{t-1}^{(3)}}^2} &=& 2\mathrm{Var}(S_{3t}) + 2\mathrm{Var}(S_{1t}) + 13\mathrm{Var}(V_{3}) + 4\mathrm{Var}(V_{2}) 
+ 20\mathrm{Var}(V_{1})\nonumber\\
&& + 4\mathrm{E}(V_{3})^2 -
16\mathrm{E}(V_{3})\mathrm{E}(V_{1}) + 4\mathrm{E}(V_{1})^2,
\quad\forall t\geq 6\\
&&\nonumber\\
\cov{{X_{t}^{(3)}}^2}{{X_{t-2}^{(3)}}^2} &=& \mathrm{Var}(S_{3t}) + \mathrm{Var}(S_{1t}) + 9\mathrm{Var}(V_{3}) + 4\mathrm{Var}(V_{2}) + 16
\mathrm{Var}(V_{1})\nonumber\\
&& + 4\mathrm{E}(V_{3})\mathrm{E}(V_{1}),
\quad\forall t\geq 7\\
&&\nonumber\\
\cov{{X_{t}^{(3)}}^2}{{X_{t-3}^{(3)}}^2} &=& \mathrm{Var}(S_{2t}) + 2\mathrm{Var}(S_{1t}) + 9\mathrm{Var}(V_{3}) + 4\mathrm{Var}(V_{2}) + 
20\mathrm{Var}(V_{1})\nonumber\\
&& + 8\mathrm{E}(V_{2})\mathrm{E}(V_{1}) + 4\mathrm{E}(V_{1})^2,
\quad\forall t\geq 8\\
&&\nonumber\\
\cov{{X_{t}^{(3)}}^2}{{X_{t-4}^{(3)}}^2} &=& \mathrm{Var}(S_{1t}) +
9\mathrm{Var}(V_{3}) + 4\mathrm{Var}(V_{2}) + 16\mathrm{Var}(V_{1}),
\quad\forall t\geq 9\\
&&\nonumber\\
\cov{{X_{t}^{(3)}}^2}{{X_{t-s}^{(3)}}^2} &=& 9\mathrm{Var}(V_{3})
+ 4\mathrm{Var}(V_{2}) + 16\mathrm{Var}(V_{1}), \quad\forall s\geq 5,
\forall t\geq s+5
\label{eq:far3}
\end{eqnarray}
\end{lemma}

\begin{lemma}
The covariance structure between the ${X_t^{(1)}}^2$ and
${X_t^{(2)}}^2$ terms is given by the following relations.
\begin{eqnarray}
\cov{{X_{t}^{(1)}}^2}{{X_{t+s}^{(2)}}^2} &=& 2(\mathrm{Var}(V_{3}) + 2\mathrm{Var}(V_{2}) + 12\mathrm{Var}(V_{1}))
, \quad\forall t\geq 3,\forall s\geq 4\label{eq:far12a}\\
&&\nonumber\\
\cov{{X_{t}^{(1)}}^2}{{X_{t+3}^{(2)}}^2} &=& \mathrm{Var}(S_{1t}) +
2\mathrm{Var}(V_{3}) + 4\mathrm{Var}(V_{2}) + 24\mathrm{Var}(V_{1}),
\quad\forall t\geq 3\\
&&\nonumber\\
\cov{{X_{t}^{(1)}}^2}{{X_{t+2}^{(2)}}^2} &=& \mathrm{Var}(S_{2t}) + 5\mathrm{Var}(S_{1t}) + 2\mathrm{Var}(V_{3}) + 4\mathrm{Var}(V_{2}) +
32\mathrm{Var}(V_{1})\nonumber\\
&& + 12\mathrm{E}(V_{2})\mathrm{E}(V_{1}) +
8\mathrm{E}(V_{1})^2, \quad\forall t\geq 3\\
&&\nonumber\\
\cov{{X_{t}^{(1)}}^2}{{X_{t+1}^{(2)}}^2} &=& \mathrm{Var}(S_{3t}) + \mathrm{Var}(S_{2t}) + 6\mathrm{Var}(S_{1t}) + 2\mathrm{Var}(V_{3}) +
4\mathrm{Var}(V_{2}) + 20\mathrm{Var}(V_{1})\nonumber\\
&& + 4\mathrm{E}(V_{3})\mathrm{E}(V_{2}) + 8\mathrm{E}(V_{3})\mathrm{E}(V_{1}
) + 8\mathrm{E}(V_{2})\mathrm{E}(V_{1}) - 4\mathrm{E}(V_{1})^2,\
\forall t\geq 3\\ 
&&\nonumber\\
\cov{{X_{t}^{(1)}}^2}{{X_{t}^{(2)}}^2} &=& \mathrm{Var}(S_{3t}) +
\mathrm{Var}(S_{2t})  + 6\mathrm{Var}(S_{1t}) + 2\mathrm{Var}(V_{3}) + 
4\mathrm{Var}(V_{2}) + 20\mathrm{Var}(V_{1})\nonumber\\
&& +
4\mathrm{E}(V_{3})\mathrm{E}(V_{2}) +
8\mathrm{E}(V_{3})\mathrm{E}(V_{1}) +
8\mathrm{E}(V_{2})\mathrm{E}(V_{1}) - 4\mathrm{E}(V_{1})^2,\ \forall
t\geq 4\\
&&\nonumber\\
\cov{{X_{t}^{(1)}}^2}{{X_{t-1}^{(2)}}^2} &=& \mathrm{Var}(S_{2t}) +
5\mathrm{Var}(S_{1t}) + 2\mathrm{Var}(V_{3}) + 4\mathrm{Var}(V_{2})  +  
32\mathrm{Var}(V_{1})\nonumber\\
&& + 12\mathrm{E}(V_{2})\mathrm{E}(V_{1}) +
8\mathrm{E}(V_{1})^2, \quad\forall t\geq 5\\
&&\nonumber\\
\cov{{X_{t}^{(1)}}^2}{{X_{t-2}^{(2)}}^2} &=& \mathrm{Var}(S_{1t}) +
2\mathrm{Var}(V_{3}) + 4\mathrm{Var}(V_{2}) + 24\mathrm{Var}(V_{1}),
\quad\forall t\geq 6\\
&&\nonumber\\
\cov{{X_{t}^{(1)}}^2}{{X_{t-s}^{(2)}}^2} &=& 2(\mathrm{Var}(V_{3}) +
2\mathrm{Var}(V_{2}) + 12\mathrm{Var}(V_{1})), \quad\forall s\geq
3,\forall t\geq s+4\label{eq:far12b}
\end{eqnarray}
\end{lemma}

\begin{lemma}
The covariance structure between the ${X_t^{(1)}}^2$ and
${X_t^{(3)}}^2$ terms is given by the following relations.
\begin{eqnarray}
\cov{{X_{t}^{(1)}}^2}{{X_{t+s}^{(3)}}^2} &=&
3\mathrm{Var}(V_{3}) + 4\mathrm{Var}(V_{2}) + 24\mathrm{Var}(V_{1}),
\quad\forall s\geq 5,\forall t\geq 3\label{eq:far13a}\\
&&\nonumber\\
\cov{{X_{t}^{(1)}}^2}{{X_{t+4}^{(3)}}^2} &=& \mathrm{Var}(S_{1t}) +
3\mathrm{Var}(V_{3}) + 4\mathrm{Var}(V_{2}) + 24\mathrm{Var}(V_{1}),
\quad\forall t\geq 3\\
&&\nonumber\\
\cov{{X_{t}^{(1)}}^2}{{X_{t+3}^{(3)}}^2} &=& \mathrm{Var}(S_{2t}) +
5\mathrm{Var}(S_{1t}) + 3\mathrm{Var}(V_{3})  + 4\mathrm{Var}(V_{2}) + 
32\mathrm{Var}(V_{1})\nonumber\\
&& + 12\mathrm{E}(V_{2})\mathrm{E}(V_{1}) +
8\mathrm{E}(V_{1})^2 , \quad\forall t\geq 3\\
&&\nonumber\\
\cov{{X_{t}^{(1)}}^2}{{X_{t+2}^{(3)}}^2} &=& \mathrm{Var}(S_{3t}) + \mathrm{Var}(S_{2t}) + 5\mathrm{Var}(S_{1t}) + 3\mathrm{Var}(V_{3}) + 
4\mathrm{Var}(V_{2})\nonumber\\
&& + 32\mathrm{Var}(V_{1}) + 4\mathrm{E}(V_{3})\mathrm{E}(V_{2}) + 12\mathrm{E}(V_{3})\mathrm{E}(V(1
))\nonumber\\
&& + 12\mathrm{E}(V_{2})\mathrm{E}(V_{1}) + 8\mathrm{E}(V_{1})^2,
\quad\forall t\geq 3\\
&&\nonumber\\
\cov{{X_{t}^{(1)}}^2}{{X_{t+1}^{(3)}}^2} &=& \mathrm{Var}(S_{3t}) + 2\mathrm{Var}(S_{1t}) + 3\mathrm{Var}(V_{3}) + 4\mathrm{Var}(V_{2}) + 
28\mathrm{Var}(V_{1})\nonumber\\
&& - 8\mathrm{E}(V_{3})\mathrm{E}(V_{1}) +
4\mathrm{E}(V_{1})^2, \quad\forall t\geq 4\\
&&\nonumber\\
\cov{{X_{t}^{(1)}}^2}{{X_{t}^{(3)}}^2}  &=& \mathrm{Var}(S_{3t}) + \mathrm{Var}(S_{2t}) + 5\mathrm{Var}(S_{1t}) + 3\mathrm{Var}(V_{3}) + 
4\mathrm{Var}(V_{2})\nonumber\\
&& + 32\mathrm{Var}(V_{1}) + 4\mathrm{E}(V_{3})\mathrm{E}(V_{2}) + 12\mathrm{E}(V_{3})\mathrm{E}(V_{1})\nonumber\\
&& + 12\mathrm{E}(V_{2})\mathrm{E}(V_{1}) + 8\mathrm{E}(V_{1})^2,
\quad\forall t\geq 5\\
&&\nonumber\\
\cov{{X_{t}^{(1)}}^2}{{X_{t-1}^{(3)}}^2}  &=& \mathrm{Var}(S_{2t}) + 5\mathrm{Var}(S_{1t}) + 3\mathrm{Var}(V_{3}) + 4\mathrm{Var}(V_{2}) +
32\mathrm{Var}(V_{1})\nonumber\\
&& + 12\mathrm{E}(V_{2})\mathrm{E}(V_{1}) +
8\mathrm{E}(V_{1})^2, \quad\forall t\geq 6\\
&&\nonumber\\
\cov{{X_{t}^{(1)}}^2}{{X_{t-2}^{(3)}}^2} &=& \mathrm{Var}(S_{1t}) +
3\mathrm{Var}(V_{3}) + 4\mathrm{Var}(V_{2}) + 24\mathrm{Var}(V_{1}),
\quad\forall t\geq 7\\
&&\nonumber\\
\cov{{X_{t}^{(1)}}^2}{{X_{t-s}^{(3)}}^2} &=& 3\mathrm{Var}(V_{3}) +
4\mathrm{Var}(V_{2}) + 24\mathrm{Var}(V_{1}), \quad\forall s\geq
3,\forall t\geq s+4 \label{eq:far13b}
\end{eqnarray}
\end{lemma}

\begin{lemma}
The covariance structure between the ${X_t^{(2)}}^2$ and
${X_t^{(3)}}^2$ terms is given by the following relations.
\begin{eqnarray}
\cov{{X_{t}^{(2)}}^2}{{X_{t+s}^{(3)}}^2} &=& 2(3\mathrm{Var}(V_{3}) + 
2\mathrm{Var}(V_{2}) + 8\mathrm{Var}(V_{1})), \quad\forall t\geq 4, 
\forall s\geq 5 \label{eq:far23a}\\
&&\nonumber\\
\cov{{X_{t}^{(2)}}^2}{{X_{t+4}^{(3)}}^2} &=& \mathrm{Var}(S_{1t}) + 
6\mathrm{Var}(V_{3}) + 4\mathrm{Var}(V_{2}) + 16
\mathrm{Var}(V_{1}), \quad\forall t\geq 4\\
&&\nonumber\\
\cov{{X_{t}^{(2)}}^2}{{X_{t+3}^{(3)}}^2} &=& \mathrm{Var}(S_{2t}) + 
2\mathrm{Var}(S_{1t}) + 6\mathrm{Var}(V_{3}) + 4
\mathrm{Var}(V_{2}) +
20\mathrm{Var}(V_{1})\nonumber\\
&& + 8\mathrm{E}(V_{2})\mathrm{E}(V_{1}) + 
4\mathrm{E}(V_{1})
^2, \quad\forall t\geq 4\\
&&\nonumber\\
\cov{{X_{t}^{(2)}}^2}{{X_{t+2}^{(3)}}^2} &=& \mathrm{Var}(S_{3t}) + 
2\mathrm{Var}(S_{1t}) + 6\mathrm{Var}(V_{3}) + 4
\mathrm{Var}(V_{2}) +
12\mathrm{Var}(V_{1})\nonumber\\
&& - 4\mathrm{E}(V_{1})^2, \quad\forall 
t\geq 4\\ 
&&\nonumber\\
\cov{{X_{t}^{(2)}}^2}{{X_{t+1}^{(3)}}^2} &=& 2\mathrm{Var}(S_{3t}) + \mathrm{Var}(S_{2t}) + 3\mathrm{Var}(S_{1t}) + 10\mathrm{Var}(V_{3})
 + 4\mathrm{Var}(V_{2})\nonumber\\
&& + 12\mathrm{Var}(V_{1}) + 4\mathrm{E}(V_{3})^2 + 8\mathrm{E}(V_{3})\mathrm{E}(V_{2}) +
 8\mathrm{E}(V_{3})\mathrm{E}(V_{1})\nonumber\\
&& +
 4\mathrm{E}(V_{2})\mathrm{E}(V_{1}) - 4\mathrm{E}(V_{1})^2,
 \quad\forall t\geq 4\\
&&\nonumber\\
\cov{{X_{t}^{(2)}}^2}{{X_{t}^{(3)}}^2} &=& 2\mathrm{Var}(S_{3t}) + \mathrm{Var}(S_{2t}) + 3\mathrm{Var}(S_{1t}) + 10\mathrm{Var}(V_{3})
 + 4\mathrm{Var}(V_{2})\nonumber\\
&& + 12\mathrm{Var}(V_{1}) + 4\mathrm{E}(V_{3})^2 +
8\mathrm{E}(V_{3})\mathrm{E}(V_{2}) + 
 8\mathrm{E}(V_{3})\mathrm{E}(V_{1})\nonumber\\
&& +
 4\mathrm{E}(V_{2})\mathrm{E}(V_{1})  - 4\mathrm{E}(V_{1})^2,
 \quad\forall t\geq 5\\
&&\nonumber\\
\cov{{X_{t}^{(2)}}^2}{{X_{t-1}^{(3)}}^2} &=& \mathrm{Var}(S_{3t}) + 2\mathrm{Var}(S_{1t}) + 6\mathrm{Var}(V_{3}) + 4\mathrm{Var}(V_{2}) + 
12\mathrm{Var}(V_{1})\nonumber\\
&& - 4\mathrm{E}(V_{1})^2, \quad\forall t\geq 6\\
&&\nonumber\\
\cov{{X_{t}^{(2)}}^2}{{X_{t-2}^{(3)}}^2} &=& \mathrm{Var}(S_{2t}) + 2\mathrm{Var}(S_{1t}) + 6\mathrm{Var}(V_{3}) + 4\mathrm{Var}(V_{2}) + 
20\mathrm{Var}(V_{1})\nonumber\\
&& + 8\mathrm{E}(V_{2})\mathrm{E}(V_{1}) +
4\mathrm{E}(V_{1})^2, \quad\forall t\geq 7\\
&&\nonumber\\
\cov{{X_{t}^{(2)}}^2}{{X_{t-3}^{(3)}}^2} &=& \mathrm{Var}(S_{1t}) +
6\mathrm{Var}(V_{3}) + 4\mathrm{Var}(V_{2}) + 16\mathrm{Var}(V_{1}),
\quad\forall t\geq 8\\
&&\nonumber\\
\cov{{X_{t}^{(2)}}^2}{{X_{t-s}^{(3)}}^2} &=& 2(3\mathrm{Var}(V_{3}) +
2\mathrm{Var}(V_{2}) + 8\mathrm{Var}(V_{1})), \quad\forall s\geq 4,
\forall t\geq s+5
\label{eq:far23b} 
\end{eqnarray}
\end{lemma}

}

\begin{proof}
All results were first derived using a simple program for the \reduce\
computer algebra system. The output
was used for further computations, and also passed through a filter
which type-set the results accurately for \LaTeX. All
calculations and computations were verified by re-coding the entire
problem in a very
different way for the \mupad\
computer algebra system. The author would be willing to e-mail the
\reduce\ and/or \mupad\ programs to any interested parties.   
\end{proof}

Note that the ``far-away'' covariances, \eqref{eq:far1}, \eqref{eq:far2},
\eqref{eq:far3}, \eqref{eq:far12a}, \eqref{eq:far12b}, \eqref{eq:far13a},
\eqref{eq:far13b}, \eqref{eq:far23a} and \eqref{eq:far23b} may be
deduced directly using Lemma \ref{lem:id} and the $n$-step
exchangeability representation theorem. Unfortunately, the other
covariances are not so amenable to such an approach.

\bibliography{bayeslin,djw}
\bibliographystyle{plain}
 
\end{document}